\documentclass{elsart}

\usepackage{graphicx}
\usepackage{ifthen}
\usepackage{xspace}

\usepackage{amsmath}

\begin{document}

\begin{frontmatter}
\title{Study of Hadronic Five-Body Decays of Charmed Mesons Involving $K_S^0$}

\date{\today}

The FOCUS Collaboration

\author[ucd]{J.~M.~Link}
\author[ucd]{P.~M.~Yager}
\author[cbpf]{J.~C.~Anjos}
\author[cbpf]{I.~Bediaga}
\author[cbpf]{C.~G\"obel}
\author[cbpf]{A.~A.~Machado}
\author[cbpf]{J.~Magnin}
\author[cbpf]{A.~Massafferri}
\author[cbpf]{J.~M.~de~Miranda}
\author[cbpf]{I.~M.~Pepe}
% ...

\address[ucd]{University of California, Davis, CA 95616} 
\address[cbpf]{Centro Brasileiro de Pesquisas F\'\i sicas, Rio de Janeiro, RJ, Brasil} 
\address[cinv]{CINVESTAV, 07000 M\'exico City, DF, Mexico} 
\address[cu]{University of Colorado, Boulder, CO 80309} 
\address[fnal]{Fermi National Accelerator Laboratory, Batavia, IL 60510} 
\address[fras]{Laboratori Nazionali di Frascati dell'INFN, Frascati, Italy I-00044}
\address[guan]{University of Guanajuato, 37150 Leon, Guanajuato, Mexico} 
\address[ui]{University of Illinois, Urbana-Champaign, IL 61801} 
\address[iu]{Indiana University, Bloomington, IN 47405} 
\address[korea]{Korea University, Seoul, Korea 136-701}
\address[korea2]{Kyungpook National University, Taegu, Korea 702-701}
\address[milan]{INFN and University of Milano, Milano, Italy} 
\address[nc]{University of North Carolina, Asheville, NC 28804} 
\address[pavia]{Dipartimento di Fisica Nucleare e Teorica and INFN, Pavia, Italy} 
\address[pr]{University of Puerto Rico, Mayaguez, PR 00681} 
\address[sc]{University of South Carolina, Columbia, SC 29208} 
\address[ut]{University of Tennessee, Knoxville, TN 37996} 
\address[vu]{Vanderbilt University, Nashville, TN 37235} 
\address[wisc]{University of Wisconsin, Madison, WI 53706}

\endnote{\small See http://www-focus.fnal.gov/authors.html for
additional author information}


\begin{abstract}
We study
the decay of $D^0$ and $D^+_s$ mesons into five-body final states including a $K_S^0$ and
report the discovery of the decay mode $D_s^+\rightarrow K_S^0 K_S^0\pi^+\pi^+\pi^-$. The
branching ratio for the new mode is {${\Gamma(D_s^+\rightarrow
K_S^0K_S^0\pi^+\pi^-\pi^+)}\over{\Gamma(D_s^+\rightarrow
K_S^0K^-\pi^+\pi^+)}$} = 0.102$\pm$0.029$\pm$0.029.  We 
also determine the branching ratio of 
{${\Gamma(D^0\rightarrow K_S^0\pi^+\pi^+\pi^-\pi^-)}\over{\Gamma(D^0\rightarrow
K_S^0\pi^+\pi^-)}$} = 0.095$\pm$0.005$\pm$0.007 
as well as an upper limit for 
{${\Gamma(D^0\rightarrow K_S^0K^-\pi^+\pi^+\pi^-)}\over{\Gamma(D^0\rightarrow
K_S^0\pi^+\pi^+\pi^-\pi^-)}$}~$<$~0.054 (90\% CL).
 An analysis of the resonant substructure for $D^0 \rightarrow K_S^0\pi^+\pi^+\pi^-\pi^-$ 
is also performed. 
 
\end{abstract}
\end{frontmatter}



More information on multibody final states in the charm sector 
is an essential ingredient for our
ability to model decay rates and to further increase our understanding of the
decay process in heavy quark systems. This is particularly important 
for the $D^+_s$ decays where a substantial part of its hadronic decay rate is still not
identified. In this paper we extend our work [1]  on four-body decays involving 
a $K^0_S$ to five-body decays involving a $K^0_S$. We have already published 
results on all charged five-body modes [2]. The FOCUS collaboration 
presents the first evidence of the
decay mode $D_s^+\rightarrow K_S^0 K_S^0\pi^+\pi^+\pi^-$, measures an inclusive
branching ratio for the mode $D^0\rightarrow K_S^0 \pi^+\pi^+\pi^-\pi^-$ relative
to $D^0\rightarrow K_S^0 \pi^+\pi^-$ and places
an upper limit on the mode $D^0\rightarrow K_S^0 K^-\pi^+\pi^+\pi^-$. Finally we
present the first resonant substructure analysis of the decay mode 
$D^0\rightarrow K_S^0 \pi^+\pi^+\pi^-\pi^-$.

The data were collected during the 1996-1997 fixed target run at Fermilab.
Bremsstrahlung of electrons and photons with an endpoint energy of approximately
300 GeV produces  photons which interact in a segmented
beryllium-oxide target to produce charmed particles. The average photon energy 
for events which satisfy our trigger is $\approx$ 180~GeV. Charged decay products
are momentum analyzed by two oppositely polarized dipole magnets. Tracking is performed
by a system of silicon vertex detectors [3] in the target region and by multi-wire
proportional chambers downstream of the interaction. Particle identification is
performed by three threshold \v{C}erenkov counters, two electromagnetic calorimeters,
a hadronic calorimeter, and two muon systems.

Five-body $D^0$ and $D_s^+$ decays are reconstructed using a candidate driven
vertex algorithm [4]. A decay vertex is formed from the reconstructed 
charged tracks. The $K_S^0$ is also reconstructed using techniques described
elsewhere [5]. The momentum
information from the $K_S^0$ and the charged tracks is used to form a
candidate $D$ momentum vector, which is intersected with other tracks to find
the production vertex. Events are selected based on several criteria. The
confidence level for the production vertex and for the charm decay vertex must be 
greater than 1$\%$. 
The reconstructed mass of the $K_S^0$ must be within four
standard deviations of the nominal $K_S^0$ mass.
The likelihood for each charged particle to be a proton,
kaon, pion, or electron based on \v{C}erenkov particle identification is used to
make additional requirements [6]. 
For pion candidates we require a loose cut that no alternative 
hypothesis is favored over the pion hypothesis by more than 6 units of log-likelihood.
In addition, for each
kaon candidate we require the negative log-likelihood kaon hypothesis, $W_K =
-2$ ln(kaon likelihood), to be favored over the corresponding pion hypothesis
$W_{\pi}$ by $W_{\pi} - W_K > 2$. We also require the distance between the
primary and secondary vertices divided by its error 
to be at least 10. Finally, in order to reduce
background due to secondary interactions of particles from the production
vertex, we require the secondary vertex to be located outside the target
material. 

\begin{figure}[bht]
\begin{center}
\includegraphics[width=12cm]{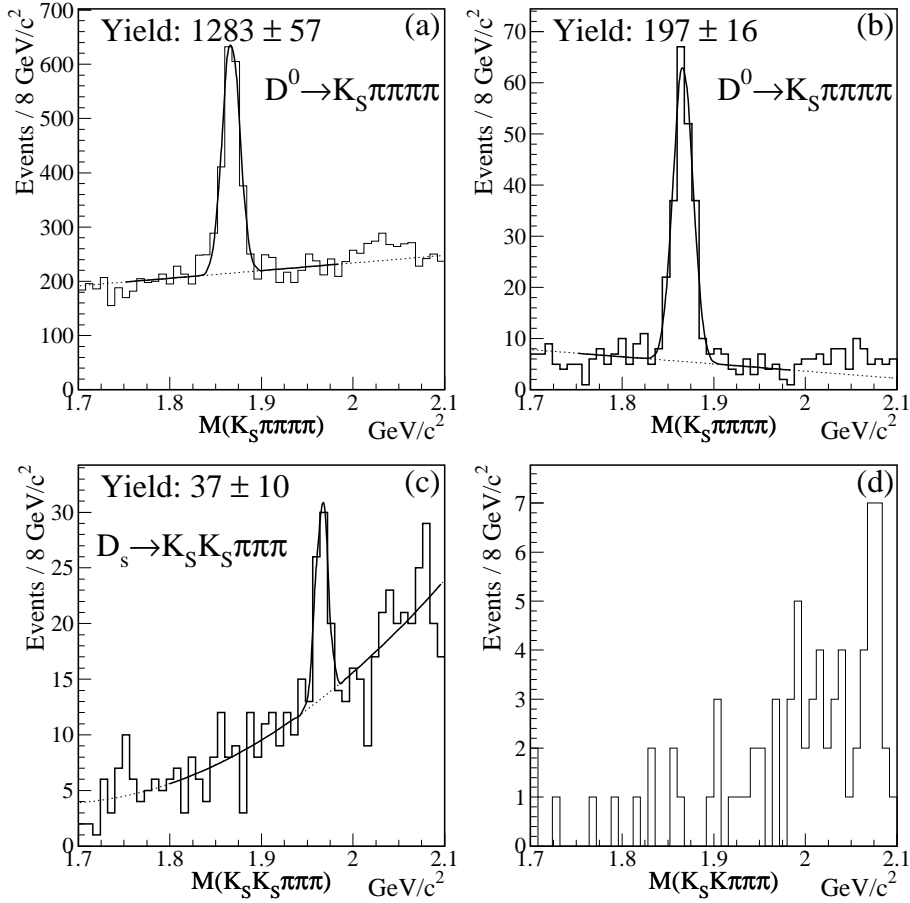}
\end{center}
\caption{Invariant mass distributions for  (a) $K_S^0 \pi^+\pi^+\pi^-\pi^-$, 
(b) $K_S^0 \pi^+\pi^+\pi^-\pi^-$  for $D^*$ tagged events,
(c) $K_S^0 K_S^0 \pi^+\pi^-\pi^{\pm}$, and 
(d) $K_S^0 K^-\pi^+\pi^-\pi^+$. The fits are 
described in the text.}
\label{Fig 1}
\end{figure}

For individual modes we apply additional analysis cuts.
Due to the large combinatoric background for $D^0\rightarrow K_S^0\pi^+\pi^+\pi^-\pi^-$,
we increase the
separation requirement of the secondary vertex from being just outside the target material
to two standard deviations from the edge of the target material. Figure 1(a) shows
the $K_S^0\pi^+\pi^+\pi^-\pi^-$ invariant mass plot for events that satisfy these
cuts. The distribution is fitted with a Gaussian for the $D^0$ signal
(1283$\pm$57 events) with the width and mass floated 
and a first degree polynomial for the background. Figure 1(b) shows
the $K_S^0\pi^+\pi^+\pi^-\pi^-$ invariant mass plot for events originating from a
$D^{*+}\rightarrow D^0\pi^+$ decay.

The $D_s^+\rightarrow K_S^0 K_S^0\pi^+\pi^+\pi^-$ mode is difficult to detect due
to the relative inefficiency of $K_S^0$ reconstruction and that most of the time only
the three pions define the secondary vertex.  
The confidence level that a pion track from the decay vertex intersects
the production vertex must be less than 2$\%$. We also require a reconstructed
$D_s^+$ momentum of greater than 25 GeV/$c$. 
Figure 1(c) shows the $K_S^0 K_S^0\pi^+\pi^+\pi^-$ mass plot for events which satisfy these
cuts. This is the first observation of this mode. We fit with a Gaussian
(37$\pm$10 events) with mass and width allowed to float and a second degree polynomial for the background.

The decay $D^0\rightarrow K_S^0 K^-\pi^+\pi^+\pi^-$ is Cabibbo suppressed, and we 
do not observe a signal in this mode. Thus we choose our analysis cuts by
maximizing the quantity $S/\sqrt{B}$, where $S$ is the fitted yield from our
Monte Carlo simulation of the mode, and $B$ is the number of background events
in the signal region from data. Based on this optimization 
we require a reconstructed $D^0$ momentum of greater than 
50 GeV/$c$. We also require the $D^0$ come from a $D^{*+}$ decay, that is 
0.142~GeV/$c^2~<M_{D^{*+}}-M_{D^0}<$0.149~GeV/$c^2$. Figure 1(d) shows the resulting 
$K_S^0 K^-\pi^+\pi^+\pi^-$ invariant mass plot. 
As there is no apparent signal we report an upper limit branching ratio.

\begin{table}[htb]
\begin{center}
\caption{Branching ratios for modes involving a $K_S^0$. All
branching ratios are inclusive of subresonant modes.}
\begin{tabular}{cc} \hline \hline
Decay Mode&Branching Ratio\\
\hline
$\frac{\Gamma(D^0\!\rightarrow K_S^0\pi^+\pi^+\pi^-\pi^-)}{\Gamma(D^0\!\rightarrow
K_S^0\pi^+\pi^-)}$&0.095$\pm$0.005$\pm$0.007\\
$\frac{\Gamma(D_s^+\!\rightarrow K_S^0K_S^0\pi^+\pi^+\pi^-)}{\Gamma(D_s^+\!\rightarrow
K_S^0 K^-\pi^+\pi^+)}$&0.102$\pm$0.029$\pm$0.029\\
$\frac{\Gamma(D^0\!\rightarrow K_S^0 K^-\pi^+\pi^+\pi^-)}{\Gamma(D^0\!\rightarrow
K_S^0\pi^+\pi^+\pi^-\pi^-)}$&$<$ 0.054 (90$\%$ C.L.)\\
\hline
\hline
\end{tabular}
\end{center}
\end{table}
We measure the branching fraction of the $D^0\rightarrow
K_S^0\pi^+\pi^+\pi^-\pi^-$ mode relative to $D^0\rightarrow
K_S^0\pi^+\pi^-$. The relative efficiency is determined by Monte
Carlo simulation. The $K_S^0\pi^+\pi^-$ and $K_S^0\pi^+\pi^+\pi^-\pi^-$
channels are produced as an incoherent mixture of 
subresonant decays based on PDG information [7] and our analysis described 
below, respectively. We measure the 
 $D_s^+\rightarrow K_S^0 K_S^0\pi^+\pi^+\pi^-$ mode 
relative to $D_s^+\rightarrow K_S^0 K^-\pi^+\pi^+$.
We test for dependency on
cut selection in both modes by individually varying each cut. The results are
shown in Table 1, and we compare our measurement of the $D^0\rightarrow
K_S^0\pi^+\pi^+\pi^-\pi^-$ branching ratio with previous measurements
in Table 2. 

\begin{table}[t]   
\begin{center}
\caption{Comparison of this measurement of $D^0\rightarrow K_S^0\pi^+\pi^+\pi^-\pi^-$ 
mode to previous measurements.}
\begin{tabular}{ccc} \hline \hline
Experiment&Events&$\frac{\Gamma (D^0 \rightarrow K_S^0\pi^+\pi^+\pi^-\pi^-)}
 {\Gamma (D^0 \rightarrow K_S^0\pi^+\pi^-)}$\\
\hline 
E831 (This Measurement)&1283&0.095$\pm$0.005$\pm$0.007\\
PDG Average[7]& &0.107$\pm$0.029\\
ARGUS[8]& 11 &0.07$\pm$0.02$\pm$0.01\\
CLEO[9]& 56 &0.149$\pm$0.026\\
E691[10]& 6 &0.18$\pm$0.07$\pm$0.04\\
\hline
\hline
\end{tabular}
\end{center}
\end{table}

We studied systematic effects due to uncertainties in the reconstruction
efficiency, in the unknown resonant substructure, and on the fitting procedure.
To determine the systematic error due to the reconstruction efficiency we follow
a procedure based on the S-factor method used by the Particle Data Group [7].
For each mode we split the data sample into four independent subsamples based on
$D$ momentum and on the period of time in which the data was collected. These
splits provide a check on the Monte Carlo simulation of charm production, of the
vertex detector (it changed during the course of the run), and on the simulation
of the detector stability. We then define the split sample variance as the
difference between the scaled variance and the statistical variance if the
former exceeds the latter. The method is described in detail in reference [11].
In addition, we split the data sample into three
independent subsamples based on the location and geometry of the $K_S^0$ decay. We
then calculate the $K_S^0$ reconstruction variance using the same procedure
described for the split sample variance. We also vary the subresonant states
in the Monte Carlo and use the variance in the branching ratios as a
contribution to the systematic error. We also determine the systematic effects
based on different fitting procedures. The branching ratios are evaluated under
various fit conditions, and the variance of the results is used as an 
additional systematic error. 
Finally, we evaluate systematic
effects from uncertainty in the absolute tracking efficiency of multi-body decays
using studies of $D^0\rightarrow K^-\pi^+\pi^+\pi^-$ and $D^0\rightarrow
K^-\pi^+$ decays. The systematic effects are then all added in
quadrature to obtain the final systematic error. 

We do not observe a signal in the decay $D^0\rightarrow K_S^0
K^-\pi^+\pi^+\pi^-$ and we calculate an upper limit for the branching ratio with
respect to $D^0\rightarrow K_S^0 \pi^+\pi^+\pi^-\pi^-$.
We evaluate the upper limit using 
the method of Rolke and Lopez [12]. We define the signal
region as being within $\pm 2\sigma$ of the nominal $D^0$ mass, and the two sideband
regions as 4-8$\sigma$ above and below the $D^0$ mass. We 
observe 3 events in the signal region and 6 events in the sidebands,
corresponding to an upper limit of 5.02 events (@90$\%$ CL). 

We study systematic effects for this channel from cut variation and resonant
substructure, and include these in our determination of the upper limit using
the method of Cousins and Highland [13]. We determine the systematic error from
cut variation by individually varying each cut, fitting the resulting distribution,
and taking the variance between each branching ratio measurement as our
systematic error. We also study systematic effects from our uncertainty in the resonant
substructure of the mode by varying the subresonant states included in the Monte Carlo
simulation, and used the variance in the resulting branching ratios as our systematic
error. These two systematic effects are then added in quadrature to give a final
relative systematic error of 26$\%$. 

We then determine the increase in our upper limit based on the equation:
\begin{displaymath}
\Delta U = \frac{1}{2} U^2 \sigma_{sys}^2 \frac{U+b-s}{U+b}
\end{displaymath}
where U is the original upper limit of events, $\sigma_{sys}$ is the percent systematic
error determined above, $b$ is the number of events observed in the sideband region, and
$s$ is the number of signal events observed. We calculate an upper limit of 5.64
events, corresponding to an upper limit for the branching ratio of:
\begin{displaymath}
 \frac{\Gamma (D^0 \rightarrow K_S^0 K^-\pi^+\pi^+\pi^-)}
 {\Gamma (D^0 \rightarrow K_S^0 \pi^+\pi^+\pi^-\pi^-)} < 0.054 \textrm{ (@90$\%$ CL}).
\end{displaymath}
 
We have studied the
resonance substructure in the decay $D^0\rightarrow K_S^0\pi^+\pi^+\pi^-\pi^-$. We use an
incoherent binned fit method [14] developed by the E687 Collaboration which assumes the final
state is an incoherent superposition of subresonant decay modes containing vector
resonances. A coherent analysis would be difficult given our limited statistics. 
For subresonant decay modes we consider the lowest mass ($K_S^0\pi^-$) and ($\pi^+\pi^-$) 
resonances, as well
as a nonresonant channel: $K^{*-}\pi^+\pi^+\pi^-$, $K_S^0\rho^0\pi^+\pi^-$,
$K^{*-}\rho^0\pi^+$ and $(K_S^0\pi^+\pi^+\pi^-\pi^-)_{NR}$. All states not explicitly
considered are assumed to be included in the nonresonant channel. 

For the resonant substructure analysis of $D^0\rightarrow K_S^0\pi^+\pi^+\pi^-\pi^-$
we place additional cuts to enhance the signal to background ratio.
We require the confidence level that a track from the decay vertex
intersects the production vertex be less than 8$\%$. We also require the $D^0$ 
to come from a $D^{*+}$ decay, that is 0.144~GeV/$c^2~<M_{D^*+}-M_{D^0}<$0.148~GeV/$c^2$, 
in order to reduce
background and distinguish between $D^0$ and $\overline{D^0}$. Fig. 1(b) shows the 
$K_S^0\pi^+\pi^+\pi^-\pi^-$ invariant mass plot for events which satisfy these cuts.
We then determine the acceptance corrected yield into each subresonant mode using a
weighting technique whereby each event is weighted by its kinematic values
in three submasses: 
($K_S^0\pi^-$), ($\pi^+\pi^-$), and ($\pi^+\pi^+$). No resonance in the 
($\pi^+\pi^+$) submass exists, but we include it in order to compute a
meaningful $\chi^2$ estimate of the fit. Eight
population bins are constructed depending on whether each of the three submasses 
falls within the expected resonance (In the case of $\pi^+\pi^+$, the
bin is split into high and low mass regions). For each Monte Carlo simulation
the bin population, $n_i$, in the eight bins is determined and a matrix,
$T_{i\alpha}$, is calculated between the generated states, $\alpha$, Monte Carlo
yields, $Y_{\alpha}$, and the eight bins $i$:
\begin{displaymath}
n_i = \sum_{\alpha}T_{i\alpha}Y_{\alpha}~~~.
\end{displaymath} 

The elements of the matrix, $T$, can be summed to give the efficiency for
each mode, $\epsilon_{\alpha}$:
\begin{displaymath}
\epsilon_{\alpha} = \sum_{i} T_{i\alpha}~~~.
\end{displaymath}

The Monte Carlo determined matrix is inverted to create a new weighting matrix
which multiplies the bin populations to produce efficiency corrected yields. The weight
includes the contributions from the four combinations we have for each event. Each
data event can then be weighted according to its values in the submass bins. Once the
weighted distributions for each of the four modes are generated, we determine the
acceptance corrected yield by fitting the distributions with a Gaussian signal and a
linear background. Using
incoherent Monte Carlo mixtures of the four subresonant modes we verify 
that our
procedure is able to correctly recover the generated mixtures of the four
modes.

\begin{table}[t]
\begin{center}
\caption{Fractions relative to the inclusive
mode for the resonance substructure 
of the $D^0\!\rightarrow K_S^0\pi^+\pi^+\pi^-\pi^-$ decay mode. 
These values are not corrected for
unseen decay modes.}
\begin{tabular}{ccc} \hline \hline
Subresonant Mode & Fraction of $K_S^0\pi^+\pi^-\pi^+\pi^-$\\ 
\hline 
$(K_S^0\pi^+\pi^+\pi^-\pi^-)_{\textrm{NR}}$& $<$~0.46~@90\% CL\\
$K^{*-}\pi^+\pi^+\pi^-$&0.17$\pm$0.28$\pm$0.02\\
$K_S^0\rho^0\pi^+\pi^-$&0.40$\pm$0.24$\pm$0.07\\
$K^{*-}\rho^0\pi^+$&0.60$\pm$0.21$\pm$0.09\\
\hline
\hline
\end{tabular}
\end{center}
\end{table}
The results for $K_S^0\pi^+\pi^+\pi^-\pi^-$ are summarized in Table 3. The four weighted
histograms with fits are shown in Fig. 2, where Fig. 2(e) is the weighted distribution
for the sum of all subresonant modes. The goodness of fit is evaluated by calculating
a $\chi^2$ for the hypothesis of consistency between the model predictions and
observed data yields in each of the 8 submass bins. The calculated $\chi^2$ is 9.7 (4
degrees of freedom), with most of the $\chi^2$ contribution resulting from a poor
Monte Carlo simulation of the $\pi^+\pi^+$ spectrum in the nonresonant channel.

\begin{figure}[t]
\begin{center}
\includegraphics[width=4.5in]{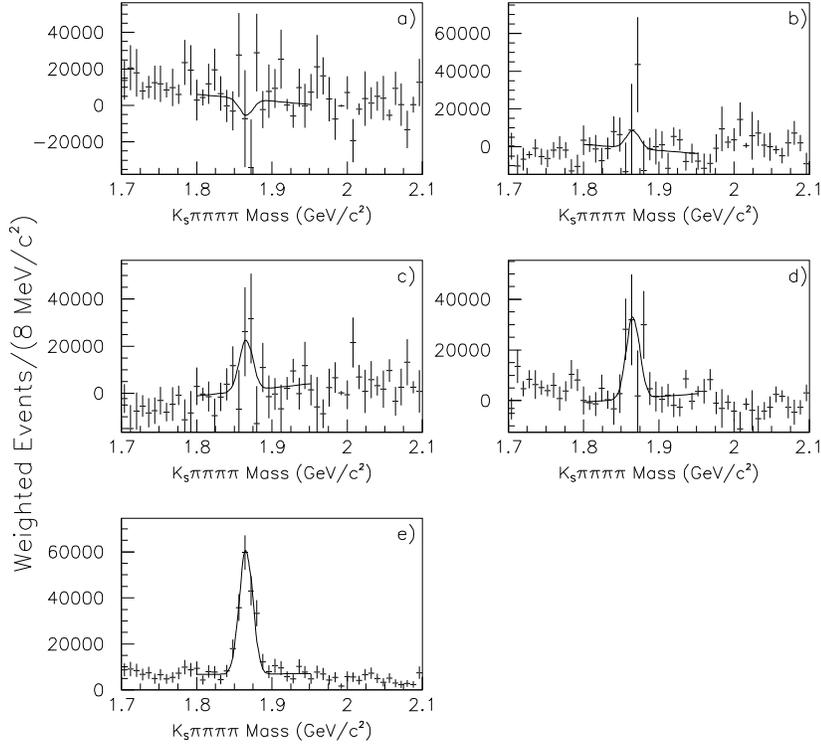}
\caption{$K_S^0\pi^+\pi^+\pi^-\pi^-$ weighted invariant mass for (a)
$(K_S^0\pi^+\pi^+\pi^-\pi^-)_{\textrm{NR}}$, (b) $K^{*-}\pi^+\pi^+\pi^-$,
(c) $K_S^0\rho^0\pi^+\pi^-$, (d) $K^{*-}\rho^0\pi^+$,
(e) Inclusive sum of all four modes.}
\label{Fig 4}
\end{center}
\end{figure}

We observe results similar to previous studies of five-body charm decays, with a small nonresonant
component and the dominant mode of the form vector-vector-pseudoscalar. Such a result
has been predicted by theoretical discussion of a vector-dominance model for heavy
flavor decays [15], which suggests that charm decays are dominated by quasi-two-body
decays in which the $W^{\pm}$ immediately hadronizes into a charged pseudoscalar,
vector or axial vector meson. 
 Results
consistent with the vector-dominance model have already been seen by FOCUS in
five-body decays [2]. Such
theoretical discussion raises the possibility that the resonant substructure for the 
decay $D^0\rightarrow K_S^0\pi^+\pi^+\pi^-\pi^-$ is dominated by the quasi-two-body decay
$K^{*-}a_1^+$. To test this hypothesis we generate Monte Carlo simulations of this
decay, assuming the $a_1^+$ has a width of 400 MeV/$c^2$ and decays 
entirely as an S-wave to $\rho^0\pi^+$, and use our subresonant analysis
procedure explained above. We observe yield fractions in each of the subresonant modes
similar to the reported fractions from the data, suggesting our results are consistent
with the decay being dominated by the $K^{*-}a_1^+$ subresonant state. 

In conclusion we have measured the relative branching ratios of many-body 
hadronic modes of $D^0$ and $D_s^+$ involving a $K_S^0$ decay and have presented the
first evidence of the decay mode $D_s^+\rightarrow K_S^0 K_S^0 \pi^+\pi^+\pi^-$. We have
also performed an analysis of the resonant substructure of the decay 
$D^0\rightarrow K_S^0\pi^+\pi^+\pi^-\pi^-$. Finally we have placed an upper limit on the
relative branching fraction of the Cabibbo suppressed decay $D^0\rightarrow K_S^0
K^-\pi^+\pi^+\pi^-$.

We acknowledge the assistance of the staffs of Fermi National
Accelerator Laboratory, the INFN of Italy, and the physics departments
of
the
collaborating institutions. This research was supported in part by the
U.~S.
National Science Foundation, the U.~S. Department of Energy, the Italian
Istituto Nazionale di Fisica Nucleare and
Ministero della Istruzione, Universit\`a e
Ricerca, the Brazilian Conselho Nacional de
Desenvolvimento Cient\'{\i}fico e Tecnol\'ogico, CONACyT-M\'exico, and
the Korea Research Foundation of
the Korean Ministry of Education.

\bibliographystyle{apsrev}
\bibliography{kshort_plb}

\begin{thebibliography}{17}
\expandafter\ifx\csname natexlab\endcsname\relax\def\natexlab#1{#1}\fi
\expandafter\ifx\csname bibnamefont\endcsname\relax
  \def\bibnamefont#1{#1}\fi
\expandafter\ifx\csname bibfnamefont\endcsname\relax
  \def\bibfnamefont#1{#1}\fi
\expandafter\ifx\csname citenamefont\endcsname\relax
  \def\citenamefont#1{#1}\fi
\expandafter\ifx\csname url\endcsname\relax
  \def\url#1{\texttt{#1}}\fi
\expandafter\ifx\csname urlprefix\endcsname\relax\def\urlprefix{URL }\fi
\providecommand{\bibinfo}[2]{#2}
\providecommand{\eprint}[2][]{\url{#2}}

\bibitem[{\citenamefont{}1}]{Link:2001gz}
\bibinfo{author}{\bibfnamefont{J.~M.} \bibnamefont{Link}}
    \bibnamefont{et~al.}
 (\bibinfo{collaboration}{FOCUS Collaboration}),
  \bibinfo{journal}{Phys. Rev. Lett.} 
  \textbf{\bibinfo{volume}{87}}, \bibinfo{pages}{162001}
  (\bibinfo{year}{2001}).
  
\bibitem[{\citenamefont{}2}]{Link:2002mm}
\bibinfo{author}{\bibfnamefont{J.~M.} \bibnamefont{Link}}
    \bibnamefont{et~al.}
 (\bibinfo{collaboration}{FOCUS Collaboration}),
  \bibinfo{journal}{Phys. Lett.}
  \textbf{\bibinfo{volume}{B 561}}, \bibinfo{pages}{225}
  (\bibinfo{year}{2003}).

\bibitem[{\citenamefont{}3}]{Link:2002zg}
\bibinfo{author}{\bibfnamefont{J.~M.} \bibnamefont{Link}}
    \bibnamefont{et~al.}
 (\bibinfo{collaboration}{FOCUS Collaboration}),
  \bibnamefont{hep-ex/0204023}.

\bibitem[{\citenamefont{}4}]{Frabetti:1992au}
\bibinfo{author}{\bibfnamefont{P.~L.} \bibnamefont{Frabetti}},
    \bibnamefont{et~al.} 
  (\bibinfo{collaboration}{E687 Collaboration}), 
  \bibinfo{journal}{Nucl. Instrum. Meth.} 
  \textbf{\bibinfo{volume}{A 320}}, \bibinfo{pages}{519}
  (\bibinfo{year}{1992}).
  
\bibitem[{\citenamefont{}5}]{Link:2001dj}
\bibinfo{author}{\bibfnamefont{J.~M.} \bibnamefont{Link}}
    \bibnamefont{et~al.}
 (\bibinfo{collaboration}{FOCUS Collaboration}),
  \bibinfo{journal}{Nucl. Instrum. Meth.}
  \textbf{\bibinfo{volume}{A 484}}, \bibinfo{pages}{174}
  (\bibinfo{year}{2001}).

\bibitem[{\citenamefont{}6}]{Link:2001pg}
\bibinfo{author}{\bibfnamefont{J.~M.} \bibnamefont{Link}}
   \bibnamefont{et~al.}
  (\bibinfo{collaboration}{FOCUS Collaboration}),
  \bibinfo{journal}{Nucl. Instrum. Meth.}
  \textbf{\bibinfo{volume}{A 484}}, \bibinfo{pages}{270}
  (\bibinfo{year}{2002}).

\bibitem[{\citenamefont{}7}]{Hagiwara:2000fs}
\bibinfo{author}{\bibfnamefont{K.} \bibnamefont{Hagiwara}} 
   \bibnamefont{et~al.} 
  (\bibinfo{collaboration}{Particle Data Group}), 
  \bibinfo{journal}{Phys. Rev.} 
  \textbf{\bibinfo{volume}{D 66}}, \bibinfo{pages}{010001}
  (\bibinfo{year}{2002}).
  
\bibitem[{\citenamefont{}8}]{Albrecht:1992tc}
\bibinfo{author}{\bibfnamefont{H.} \bibnamefont{Albrecht}} 
   \bibnamefont{et~al.} 
  (\bibinfo{collaboration}{ARGUS Collaboration}), 
  \bibinfo{journal}{Z. Phys.} 
  \textbf{\bibinfo{volume}{C 56}}, \bibinfo{pages}{7}
  (\bibinfo{year}{1992}).
  
\bibitem[{\citenamefont{}9}]{Ammar:1991au}
\bibinfo{author}{\bibfnamefont{R.} \bibnamefont{Ammar}} 
   \bibnamefont{et~al.} 
  (\bibinfo{collaboration}{CLEO Collaboration}), 
  \bibinfo{journal}{Phys. Rev.} 
  \textbf{\bibinfo{volume}{D 44}}, \bibinfo{pages}{3383}
  (\bibinfo{year}{1991}).
  
\bibitem[{\citenamefont{}10}]{Anjos:1990fv}
\bibinfo{author}{\bibfnamefont{J.~C.} \bibnamefont{Anjos}} 
   \bibnamefont{et~al.} 
  \bibinfo{journal}{Phys. Rev.} 
  \textbf{\bibinfo{volume}{D 42}}, \bibinfo{pages}{2414}
  (\bibinfo{year}{1990}).
  
\bibitem[{\citenamefont{}11}]{Link:2003al}
\bibinfo{author}{\bibfnamefont{J.~M.} \bibnamefont{Link}}
    \bibnamefont{et~al.}
 (\bibinfo{collaboration}{FOCUS Collaboration}),
  \bibinfo{journal}{Phys. Lett.}
  \textbf{\bibinfo{volume}{B 555}}, \bibinfo{pages}{167}
  (\bibinfo{year}{2003}).

\bibitem[{\citenamefont{}12}]{Rolke:2000ij}
\bibinfo{author}{\bibfnamefont{W.~A.} \bibnamefont{Rolke}
\bibnamefont{and}
\bibfnamefont{A.~M.} \bibnamefont{Lopez}}, 
  \bibinfo{journal}{Nucl. Instrum. Methods} 
  \textbf{\bibinfo{volume}{A 458}}, \bibinfo{pages}{745}
  (\bibinfo{year}{2001}).

\bibitem[{\citenamefont{}13}]{Cousins:1992qz}
\bibinfo{author}{\bibfnamefont{R.~D.} \bibnamefont{Cousins}
\bibnamefont{and}
\bibfnamefont{V.~L.} \bibnamefont{Highland}}, 
  \bibinfo{journal}{Nucl. Instrum. Methods} 
  \textbf{\bibinfo{volume}{A 320}}, \bibinfo{pages}{331}
  (\bibinfo{year}{1992}).
  
\bibitem[{\citenamefont{}14}]{Frabetti:1995al}
\bibinfo{author}{\bibfnamefont{P.~L.} \bibnamefont{Frabetti}}
    \bibnamefont{et~al.}
 (\bibinfo{collaboration}{E687 Collaboration}),
  \bibinfo{journal}{Phys. Lett.}
  \textbf{\bibinfo{volume}{B 354}}, \bibinfo{pages}{486}
  (\bibinfo{year}{1995}).

\bibitem[{\citenamefont{}15}]{Lipkin:2000gz}
\bibinfo{author}{\bibfnamefont{H.~J.} \bibnamefont{Lipkin}},
  \bibinfo{journal}{Phys. Lett.}
  \textbf{\bibinfo{volume}{B 515}}, \bibinfo{pages}{81}
  (\bibinfo{year}{2001}).

\end{thebibliography}

\end{document}